\documentclass[
preprintnumbers,
prd,
a4paper,
showpacs,
amsmath,
amssymb,
]{revtex4}
\usepackage{amsmath,amssymb}
\usepackage{graphicx}
\begin{document}
\title{Lightlike limit of the boosted Kerr black holes 
in higher-dimensional spacetimes}
\author{Hirotaka Yoshino}
\email{yoshino@gravity.phys.nagoya-u.ac.jp}
\affiliation{Department of Physics, Graduate School of Science, Nagoya
University, Chikusa, Nagoya 464-8602, Japan}
\preprint{DPNU-04-22}
\date{December 15, 2004}
\begin{abstract}
Deriving the gravitational field of the high-energy 
objects is an important theme because the black holes might be
produced in particle collisions in the brane world scenario. 
In this paper, we develop a method for boosting the metric 
of the stationary objects and taking the lightlike limit
in higher-dimensional spacetimes, using the Kerr black hole as one example. 
We analytically construct the metric of lightlike Kerr black holes and find that 
this method contains no ambiguity in contrast to the four-dimensional case.
We discuss the possible implications of our results in the brane world context.  
\end{abstract}
\pacs{04.50.+h, 04.20.Jb, 04.20.Cv, 11.10.Kk}
\maketitle

\section{Introduction}

Several scenarios in which the quantum gravity scale or the string
scale might be $O(\textrm{TeV})$ have been proposed~\cite{ref1}. 
In these scenarios, our space is the 3-brane in large extra
dimensions. These brane world scenarios attract
much attention because of their great significance, namely, the
possibility of observing the quantum gravity effects in future-planned
accelerators.  
In the far above regime of the Planck energy, the characteristic
scales of particle interactions are given by the gravitational radius of
the system and the classical description is expected to become valid. 
In this energy scale, the black hole production in the high-energy
particle collision occurs. The phenomenology of the black hole production
in accelerators was first discussed in~\cite{ref2}, and subsequently a great amount of
papers concerning this theme have appeared.

One of the necessary investigations is to analyze the black hole formation
in the high-energy particle system using the higher-dimensional general relativity. 
In the author's previous paper with Nambu~\cite{YN03},
we numerically calculated the apparent horizon in the grazing
collision of high-energy particles. Because the apparent horizon formation
is the sufficient condition for the black hole formation, our analysis 
provides the lowerbound of the cross section of the black hole production. 
In this analysis, we used the metric of the massless point particle,
which is obtained by boosting the Schwarzschild black hole to the speed of light,
fixing the energy $p$~\cite{AS71}.   However, there are various
characters of the particle that might affect its 
gravitational field, namely, the electric charge, the color charge,
the spin, the wave feature (which was recently pointed out
in~\cite{GR04}), and the length of the string. 
Because one should be able to obtain the metric of high-energy 
objects to investigate these effects,  the development of
such methods is important.

In this paper, we would like to propose the (probably) easiest method
for deriving the metric of the objects which is boosted to the speed of light. 
Because using a simple example is useful for this purpose, we consider the lightlike limit
of the boosted Kerr black holes in higher-dimensional spacetimes.
The higher-dimensional Kerr black hole solution was given by Myers and Perry~\cite{MP86}. 
Although the higher-dimensional Kerr black hole can have more than one
Kerr parameter, we restrict our attention to the 
one with only one Kerr parameter $a$ for simplicity.  
In this case, the singularity is a one-dimensional rotating ring. 
We consider two cases: the boost in the parallel
direction to the spin (i.e., the velocity vector is orthogonal
to the plane with the ring singularity) and the boost in the transverse direction to the spin
(i.e., the velocity vector is parallel to the plane with the ring singularity). 
In the boost, we fix the energy $p$ and the Kerr parameter $a$.

In the four-dimensional case, the lightlike limit of the boosted
Kerr black hole fixing the energy $p$ and the Kerr parameter $a$
has attracted a lot of attentions. 
Loust\'{o} and S\'{a}nchez considered the lightlike boost of the
Kerr black hole in the parallel direction to the spin~\cite{LS92}. They just boosted
the metric and succeeded in taking the lightlike limit.
But because they expanded the metric in the calculation,
it is applicable only at the distant region from the black hole and
the physical interpretation is difficult.  
Further, their method contained ambiguity as pointed out
in~\cite{BN, BH03}. Balasin and Nachbagauer proposed an
alternative method~\cite{BN}. They calculated the distributional energy-momentum
tensor of the Kerr black hole and boosted it to the speed of light
in arbitrary directions. 
Then they solved the Einstein's equation with this boosted energy-momentum tensor. 
Barrab\`{e}s and Hogan proposed a somewhat easier method~\cite{BH03}, in which they 
boost the Riemann tensor of the Kerr black hole to the speed of light 
and construct the spacetime so that it has the same curvature as the
boosted one. There is also a related work by Burinskii and Magli~\cite{BM00}.

We choose the simplest method here: we just boost the
metric without considering the energy-momentum tensor and the
Riemann tensor. In this sense, our method is 
similar to that of Loust\'{o} and S\'{a}nchez, but we do not expand
the metric. Using this method, we can analytically
derive the lightlike Kerr metric in the higher-dimensional spacetimes.
Although we reconfirm the ambiguity of this method in the four-dimensional case, 
we find that there is no such ambiguity in the higher-dimensional cases. 
Our analysis indicates that our method would be applicable
for a wide class of solutions and shows the condition for deriving the analytic 
metric. It also indicates that we would be able to 
calculate the lightlike metric in the general cases at least numerically.

Although our motivation is mainly mathematical, we also would like
to point out two possible implications of our results. First,
the Kerr black hole might provide the model of the
gravitational field of a particle with spin. Although the spin angular momentum
is different from the orbital angular momentum, there are several discussions
about the relation between the spinning particle and the Kerr black hole~\cite{ref3}.
Second, in the string theory, there are several discussions that
some black hole solutions should be regarded as the elementary particles~\cite{ref4},
or the Kerr black hole might be a classical model for the gravitational field 
of some string~\cite{Nis95, NR02}. 
In the last section of this paper, we give brief discussions about
the implications of our results from the viewpoint of
these contexts and the brane world scenarios.

This paper is organized as follows. In Sec. II, we analyze the boost
of the Kerr black hole in the parallel direction to the spin. 
We propose the general strategy to obtain the lightlike limit by
just boosting the metric and construct the analytic metric of the lightlike Kerr black hole.
Then we show that there is no ambiguity in this method in the higher-dimensional
cases and discuss the properties of
the resulting metric.   
In Sec. III, we repeat the similar analysis and discussions
in the case of the boost in the transverse direction to the spin. 
Sec. IV is devoted to summary and discussion. 
The discussion includes the advantage and the limitation in our method in general cases.
We also discuss the possible implications of our results, namely,
the effect of spin and the string length on the black hole formation
in the collision of the high-energy particles.

\section{Lightlike boost in the parallel direction to the spin}

In this section, we consider the lightlike limit of the Kerr black hole with
one Kerr parameter $a$
that is boosted in the parallel direction
to the spin for total spacetime dimension number $D\ge 5$. 
We introduce the Kerr-Schild coordinate $(\bar{t}, \bar{x}_1, \bar{x}_2, \bar{z}, \bar{y}_i)$ 
in which the metric becomes
\begin{multline}
ds^2=-d\bar{t}^2+d\bar{x}_1^2+d\bar{x}_2^2+d\bar{z}^2+\sum_{i}d\bar{y}_i^2
+\frac{\mu\bar{r}^{7-D}}{\bar{r}^4+a^2\left(\bar{z}^2+Y^2\right)}\\
\times
\left[d\bar{t}
+\frac{\bar{r}}{\bar{r}^2+a^2}\left(\bar{x}_1d\bar{x}_1+\bar{x}_2d\bar{x}_2\right)
+\frac{a}{\bar{r}^2+a^2}\left(\bar{x}_1d\bar{x}_2-\bar{x}_2d\bar{x}_1\right)
+\frac{\bar{z}d\bar{z}+\sum_i\bar{y}_id\bar{y}_i}{\bar{r}}\right]^2,
\end{multline}
where $i=1,...,D-4$, and $Y$ and $\bar{r}$ are defined via
\begin{align}
&~~~~~~~~Y^2=\sum_{i} \bar{y}_i^2,\\
&\frac{X^2}{\bar{r}^2+a^2}+\frac{\bar{z}^2+Y^2}{\bar{r}^2}=1, 
\label{parallel_rbar}
\end{align}
with $X^2=\bar{x}_1^2+\bar{x}_2^2$.
The total mass (the Arnowitt-Deser-Misner mass) $M$ and 
the angular momentum $J$ are given by
\begin{align}
 M&=\frac{(D-2)\Omega_{D-2}}{16\pi G_D}\mu,\\
 J&=\frac{-2}{D-2}Ma,
\end{align}
where $G_D$ and $\Omega_{D-2}$ denote the gravitational constant of the $D$-dimensional
spacetime and the $(D-2)$-dimensional area of the unit sphere, respectively.
The singularity is rotating in the $(\bar{x}_1, \bar{x}_2)$-plane. In the boost, 
we fix the energy $p=\gamma M$ and the Kerr parameter $a$. Here, $\gamma$ is
the usual $\gamma$ factor given by $\gamma\equiv 1/\sqrt{1-v^2}$ 
with the velocity of the Kerr black hole $v$. For convenience,
we introduce $P=\gamma \mu$. Now we consider the Lorentz boost
in the $\bar{z}$ direction:
\begin{align}
\bar{t}&=\gamma(t-vz),\\
\bar{z}&=\gamma(-vt+z),\\
\bar{x}_j&=x_j,\\
\bar{y}_i&=y_i,
\end{align}
where $j=1,2$. The boosted metric becomes
\begin{equation}
ds^2=ds_F^2+
\frac{\mu\bar{r}^{7-D}}{\bar{r}^4+a^2(\bar{z}^2+Y^2)}
\left(d\bar{t}^2+\frac{\bar{z}^2d\bar{z}^2}{\bar{r}^2}-\frac{2\bar{z}}{\bar{r}}
d\bar{t}d\bar{z}\right)+O(\gamma^{-1}),
\end{equation}
where $ds_F^2=-dt^2+\sum_jdx_j^2+dz^2+\sum_idy_i^2$.
In the limit $v\to 1$, this is reduced to
\begin{align}
&~~~~~~ds^2=ds_F^2+\lim_{v\to 1}f\left(\gamma(z-vt), X, Y\right)(dz-dt)^2,
\label{parallel_limit_f}\\
&f\left(\gamma(z-vt), X, Y\right)=\frac{\gamma P \bar{r}^{7-D}}{\bar{r}^4+a^2(\bar{z}^2+Y^2)}
\left(1+\frac{\bar{z}^2}{\bar{r}^2}-\frac{2\bar{z}}{\bar{r}}\right).
\label{parallel_f}
\end{align}
The strategy for obtaining this limit is as follows. We find the 
primitive function $\tilde{\Phi}$ such that
\begin{equation}
\frac{\partial}{\partial z}\tilde{\Phi}\left(\gamma(z-vt), X, Y\right)=
f\left(\gamma(z-vt), X, Y\right). 
\end{equation}
Then the limit can be obtained by
\begin{equation}
\lim_{v\to 1}f=\frac{\partial}{\partial z}\left\{\lim_{v\to 1}\tilde{\Phi}
\left(\gamma(z-vt), X, Y\right)\right\}.
\end{equation}
As we will see later, $\tilde{\Phi}$ is reduced to
\begin{equation}
\lim_{v\to 1}\tilde{\Phi}\left(\gamma(z-vt), X, Y\right)=(1/2)\textrm{sgn}(z-t)\Phi(X,Y)
+\bar{\Phi}\left(\gamma(z-t),X,Y\right),
\label{para_primitive}
\end{equation}
where $\textrm{sgn}(z-t)$ is determined by 
$\textrm{sgn}(z-t)\equiv 2\theta(z-t)-1$ with the Heviside step function $\theta(z-t)$. 
Hence we obtain
\begin{equation}
\lim_{v\to 1}f=\Phi(X,Y)\delta(z-t)
+\frac{\partial}{\partial z}\bar{\Phi}\left(\gamma(z-t),X,Y\right).
\end{equation}
The key point for writing down the analytic form of the lightlike metric
is whether we can obtain this primitive function $\tilde{\Phi}$ or not. 
Note that $\Phi(X,Y)$ comes from the even part of $f$ with respect to $\bar{z}$,
and the odd part of $f$ contributes only to $\bar{\Phi}(\gamma(z-t),X,Y)$. Because we
can show that the primitive function
of the odd part becomes a finite value independent of $z$ 
in the limit $v\to 1$ for all $D\ge 4$, 
it does not affect the result. Hence we omit the term $2\bar{z}/\bar{r}$ in eq.~\eqref{parallel_f} hereafter.

Now we construct the lightlike Kerr metric by finding the primitive function $\tilde{\Phi}$
in the higher-dimensional cases. 
Using eq.~\eqref{parallel_rbar}, we find the following relations: 
\begin{align}
&\frac{\partial\bar{r}}{\partial z}=\gamma\frac{\bar{r}^2+a^2}{\bar{r}}
\left(\bar{r}^2+a^2-\frac{X^2a^2}{\bar{r}^2+a^2}\right)^{-1}\gamma(z-vt),
\\
&\gamma(z-vt)=\textrm{sgn}(z-vt)\sqrt{\bar{r}^2-X^2-Y^2+\frac{a^2X^2}{\bar{r}^2+a^2}}.
\end{align}
Using these relations, we derive
\begin{equation}
f=\lim_{v\to 1}\textrm{sgn}(z-vt)
\frac{P\bar{r}^{6-D}}{\sqrt{(\bar{r}^2+a^2)(\bar{r}^2-\alpha)(\bar{r}^2-\beta)}}
\left(2-\frac{X^2}{\bar{r}^2+a^2}-\frac{Y^2}{\bar{r}^2}\right)\frac{\partial\bar{r}}{\partial z},
\end{equation}
where $\alpha$ and $\beta$ are given by
\begin{align}
\alpha&=\frac12\left[X^2+Y^2-a^2+\sqrt{(X^2+Y^2-a^2)^2+4a^2Y^2}\right],\\
\beta&=\frac12\left[X^2+Y^2-a^2-\sqrt{(X^2+Y^2-a^2)^2+4a^2Y^2}\right].
\end{align}
Now we introduce $R=\bar{r}^2$. Because $\bar{z}=0$ corresponds to $R=\alpha$,
we derive
\begin{equation}
\tilde{\Phi}/P=(1/2)\textrm{sgn}(z-vt)\left[2I^{(D)}_1(R)-X^2I^{(D)}_2(R)-Y^2I^{(D)}_3(R)\right],
\end{equation}
where
\begin{align}
&I^{(D)}_1(R)=\int_{\alpha}^{R}\frac{R^{(5-D)/2}}{\sqrt{(R+a^2)(R-\alpha)(R-\beta)}}dR,
\label{para_I1}\\
&I^{(D)}_2(R)=\int_{\alpha}^{R}\frac{R^{(5-D)/2}}{(R+a^2)\sqrt{(R+a^2)(R-\alpha)(R-\beta)}}dR,
\label{para_I2}\\
&I^{(D)}_3(R)=\int_{\alpha}^{R}\frac{R^{(3-D)/2}}{\sqrt{(R+a^2)(R-\alpha)(R-\beta)}}dR.
\label{para_I3}
\end{align}
Because taking the limit $v\to 1$ for $t\neq z$ corresponds to $R\to \infty$
and each integral is finite at $R=\infty$ for $D\ge 5$,
we derive
\begin{align}
&\Phi(X,Y)=2I^{(D)}_1(\infty)-X^2I^{(D)}_2(\infty)-Y^2I^{(D)}_3(\infty),
\label{parallel_F}\\
&\bar{\Phi}\left(\gamma(z-t),X,Y\right)=0.
\end{align}
The function $\Phi(X,Y)$ is expressed by the elliptic integrals, although
we do not show explicitly here because it is quite complicated.
The primitive functions $I_i^{(D)}(R)$ necessary for this calculation
are given in Appendix A. In summary, the metric of the lightlike Kerr black hole
has the following form: 
\begin{equation}
ds^2=ds_F^2+\Phi(X, Y)\delta(z-t)(dz-dt)^2.
\end{equation}
The spacetime is flat except at $z=t$. The delta function indicates
that the two coordinates in the regions $z>t$ and $z<t$ 
are discontinuously connected and
there is the distributional Riemann tensor at $z=t$, which
we call the gravitational shock wave. The function $\Phi(X,Y)$ characterizes
the structure of this shock wave and we call it the potential of the shock wave
hereafter.

Before looking at the behavior of the potential $\Phi$ for $D\ge 5$, we would like
to see what occurs in our method in the $D=4$ case.
In this case, we should be careful in the following two points. First,
there is no $y_i$ direction and we need not consider $I_3^{(4)}$. 
Because of this absence of $y_i$, there are two possible cases for 
$\bar{z}$=0: the $R=X^2-a^2$ case and the $R=0$ case.
Depending on these choices, we should adopt $[X^2-a^2, R]$ and $[0, R]$ 
as the integral regions of $I_i^{(4)}$, respectively. It turns out later that
they lead to different results, which
 represent the solutions in the $X>|a|$ region and in the $X<|a|$ region, respectively.
Second, because $I_1^{(4)}$ logarithmically diverges at $R\to\infty$, 
we should consider the $\bar{\Phi}$ part of the primitive function~\eqref{para_primitive}. 
If we take the choice such that $\bar{z}=0$ corresponds to $R=X^2-a^2$, 
$I_1^{(4)}$ and $I_2^{(4)}$ are given by
\begin{align}
I_1^{(4)}(R)&=\log\left|2R+2a^2-X^2+2\sqrt{(R+a^2)(R+a^2-X^2)}\right|-\log X^2
\nonumber \\
&\to \log\left|4\gamma^2(z-t)^2\right|-\log X^2~~(v\to 1),
\label{I1(4)_1}\\
I_2^{(4)}(R)&=\frac{2}{X^2}\sqrt{\frac{R+a^2-X^2}{R+a^2}}\to 
\frac{2}{X^2}~~(v\to 1), 
\end{align}
and we find
\begin{equation}
\lim_{v\to 1}f=2P\left[\frac{1}{|z-t|}+(-2\log |X|-1)\delta(z-t)\right].
\label{parallel_4D_result1}
\end{equation}
On the other hand, adopting the other choice in which $\bar{z}=0$ is equivalent
to $R=0$, $I_1^{(4)}$ and $I_2^{(4)}$ become
\begin{align}
I_1^{(4)}(R)&=\log\left|2R+2a^2-X^2+2\sqrt{(R+a^2)(R+a^2-X^2)}\right|
-\log\left|2a^2-X^2+2\sqrt{a^2(a^2-X^2)}\right|\nonumber\\
&\to \log\left|4\gamma^2(z-t)^2\right|
-\log\left|2a^2-X^2+2\sqrt{a^2(a^2-X^2)}\right|~~(v\to 1),\\
I_2^{(4)}(R)&=\frac{2}{X^2}\left(\sqrt{\frac{R+a^2-X^2}{R+a^2}}-
\sqrt{\frac{a^2-X^2}{a^2}}\right)
\to \frac{2}{X^2}\left(1-
\sqrt{\frac{a^2-X^2}{a^2}}\right)~~(v\to 1),
\end{align}
which lead to
\begin{equation}
\lim_{v\to 1}f=2P\left\{\frac{1}{|z-t|}+
\left[-2\log\left(|a|+\sqrt{a^2-X^2}\right)-1+\frac{\sqrt{a^2-X^2}}{|a|}\right]\delta(z-t)
\right\}. 
\label{parallel_4D_result2}
\end{equation}
The latter result~\eqref{parallel_4D_result2} is defined only in the region $|X|<|a|$.
Because the former result~\eqref{parallel_4D_result1} has the same value
at $|X|=|a|$ as that of eq.~\eqref{parallel_4D_result2}, $\lim_{v\to 1}f$ 
can be extended to the region $|X|>|a|$ using eq.~\eqref{parallel_4D_result1}. 
Since we can eliminate the $1/|z-t|$ term using the proper coordinate
transformation, we find that this result is the same as the previous result of~\cite{BN}
using the relation $P=2G_4p$. 

At this point, we should point out the ambiguity of this method for $D=4$. 
For example, we can consider the following form instead of eq.~\eqref{I1(4)_1}:
\begin{equation}
I_1^{(4)}(R)=\log\left|\left[2R+2a^2-X^2+2\sqrt{(R+a^2)(R+a^2-X^2)}\right]/e^{g(X)}\right|
+g(X)-
\log X^2
\end{equation}
with arbitrary function $g(X)$. This leads to 
\begin{equation}
\lim_{v\to 1}f=2P\left[\frac{1}{|z-t|}+\left(g(X)-2\log |X|-1\right)\delta(z-t)\right].
\end{equation}
instead of eq.~\eqref{parallel_4D_result1}. 
This ambiguity is the reason why the previous authors have developed the
various methods other than just boosting the metric~\cite{BN, BH03,BM00}. 
In fact, there would not exist the method to determine this factor $g(X)$
rigorously without checking the consistency with the boosted energy-momentum tensor
or the boosted Riemann tensor. 
Hence the above calculation is just the formal derivation
of the lightlike Kerr black hole metric. However, 
we would like to point out that there
is no ambiguity for the higher-dimensional cases, because the primitive function 
$\tilde{\Phi}$ is finite in the limit $v\to 1$. Hence in the higher-dimensional cases,
it is sufficient to consider the boost of the metric and
we need not calculate the energy-momentum tensor or the Riemann tensor
of this system. 

\begin{figure}[tb]
\centering
{\includegraphics[width=0.35\textwidth]{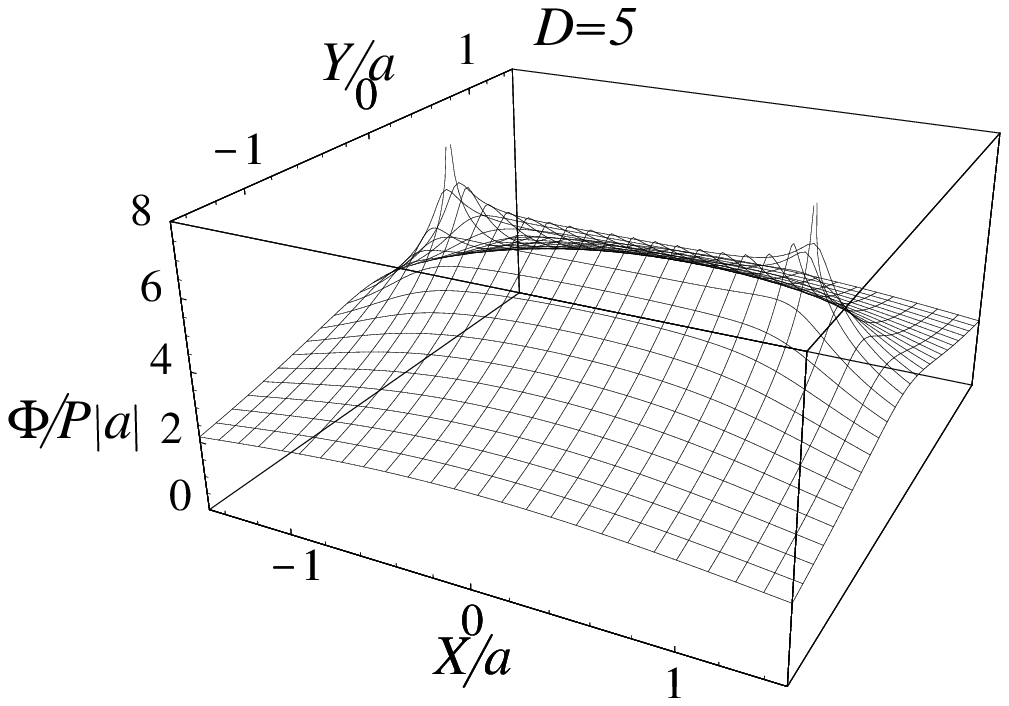}\hspace{5mm}
\includegraphics[width=0.35\textwidth]{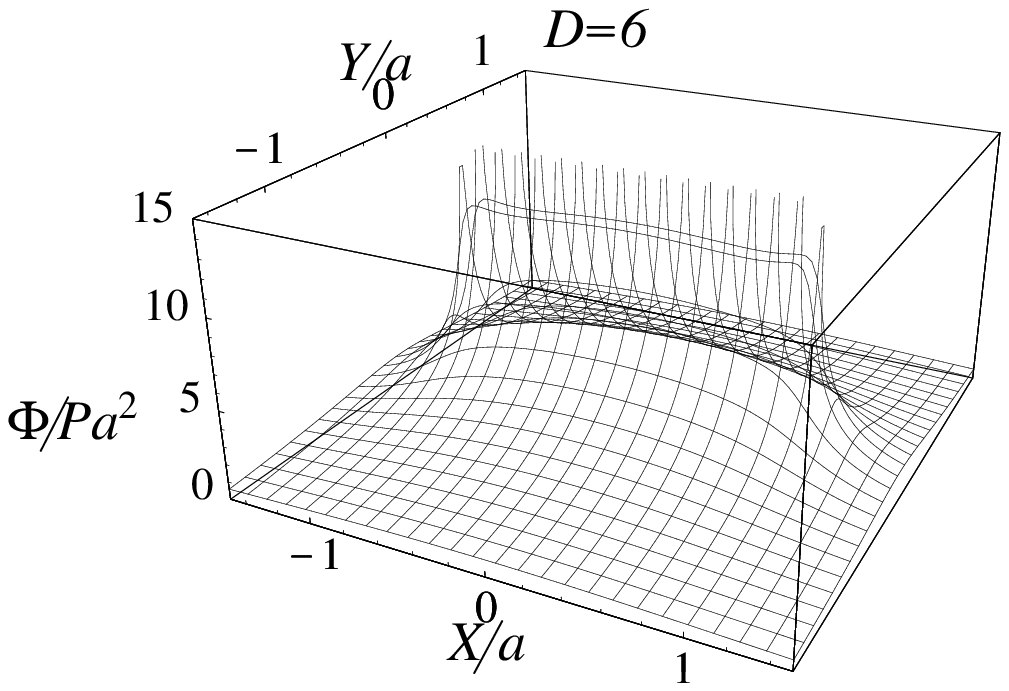}\\
\vspace{5mm}
\includegraphics[width=0.35\textwidth]{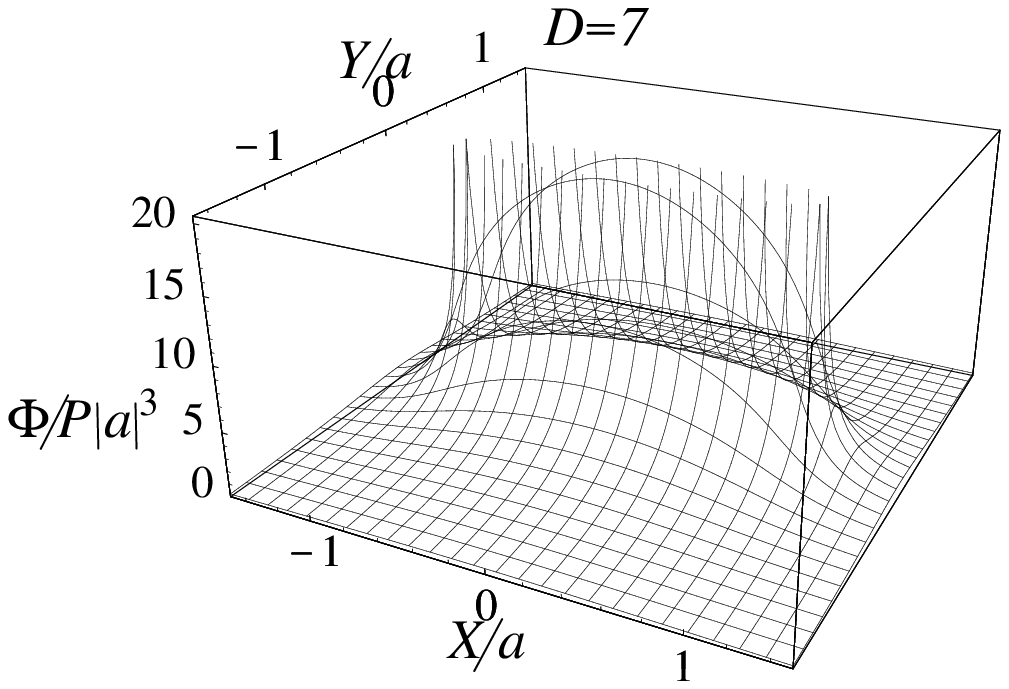}\hspace{5mm}
\includegraphics[width=0.35\textwidth]{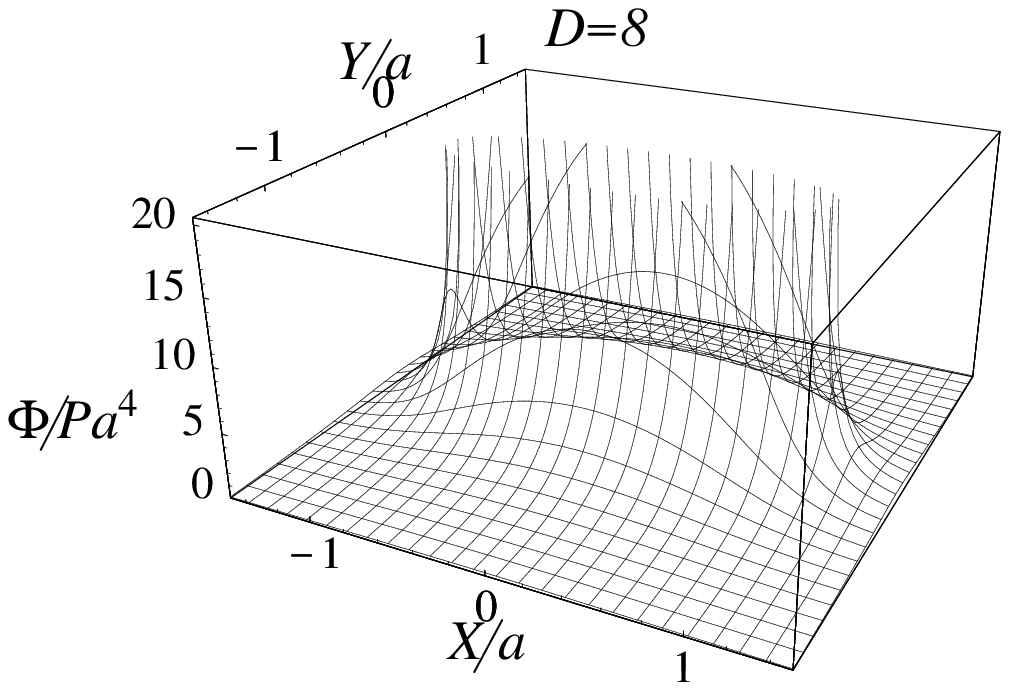}}
\caption{The behavior of the function $\Phi$ for $D=5,...,8$
in the parallel case. The unit of $X$
and $Y$ is $a$, and the unit of $\Phi$ is $P|a|^{D-4}$. In the
case of $D=5$, $\Phi$ diverges at $Y=0, X=\pm a$, which indicates that the 
singularity is a one-dimensional ring. In other cases, $\Phi$ diverges
at $Y=0, -|a|<X<|a|$. This implies that the singularity is a two-dimensional disk.}
\end{figure}

Now we discuss the properties of the potential $\Phi(X, Y)$
in the higher-dimensional cases.
Figure 1 shows the behavior of $\Phi(X,Y)$ for $D=5,...,8$. 
In the $D=5$ case, the potential $\Phi$ logarithmically diverges
at $X=\pm a$. Hence the shock wave contains a one-dimensional 
ring-shaped singularity.  
 For $D\ge 6$,
the potential $\Phi$ diverges at $Y=0, -|a|\le X\le |a|$. Hence the singularity
is a two-dimensional disk in these cases. 
This is interpreted as follows: For $D=5$, 
the black hole horizon does not exist in the resulting spacetime because
we fixed $a$ in the boosting. The singularity becomes naked and
the potential $\Phi$ diverges at the singularity. While for $D\ge 6$, 
the horizon exists for arbitrary $M$ and $a$. 
Hence the resulting gravitational shock wave contains an extremely
oblate horizon, on which the potential $\Phi$ diverges. 
The function $\Phi$ behaves like
$\log Y$ for $D=6$ and $1/Y^{D-6}$ for $D\ge 7$. One can understand this
potential behavior using the Newtonian analogy. The Newtonian 
potential of $n$-dimensional matter distribution in $N$-dimensional
space is given by $\sim 1/r^{N-n-2}$ for $N-n>2$ and $\sim -\log r$
for $N-n=2$. Because the shock wave is 
$(D-2)$ dimensional, the behavior of the potential $\Phi$
and the dimensionality of the singularity have the same relation 
as the Newtonian gravity.
Hence, we see that these results are consistent with our intuition
and we expect that also in general cases we
can guess the behavior of $\Phi(X,Y)$ to a certain extent,
even if the calculation of the primitive function of $f$ cannot be proceeded.

\section{Lightlike boost in the transverse direction of the spin}

In this section, we consider the lightlike limit of the boosted Kerr black hole in the transverse
direction to the spin. In this case, we use a coordinate system
$(\bar{t}, \bar{x}, \bar{y}, \bar{z}_i)$ in which the metric of
the Kerr black hole becomes
\begin{equation}
ds^2=-d\bar{t}^2+d\bar{x}^2+d\bar{y}^2+\sum_id\bar{z}_i^2
+\frac{\mu\bar{r}^{7-D}}{\bar{r}^4+a^2Z^2}
\left[d\bar{t} 
+\frac{\bar{r}\left(\bar{x}d\bar{x}+\bar{y}d\bar{y}\right)}{\bar{r}^2+a^2}
+\frac{a\left(\bar{x}d\bar{y}-\bar{y}d\bar{x}\right)}{\bar{r}^2+a^2}
+\frac{\sum_i\bar{z}_id\bar{z}_i}{\bar{r}}\right]^2,
\end{equation}
where $i=1,...,D-3$, and $Z$ and $\bar{r}$ are defined by
\begin{align}
&~~~~Z^2=\sum_i \bar{z}_i^2,\\
&\frac{\bar{x}^2+\bar{y}^2}{\bar{r}^2+a^2}+\frac{Z^2}{\bar{r}^2}=1.
\label{transverse_rbar}
\end{align}
The singularity is rotating in the $(\bar{x}, \bar{y})$-plane and we will consider
the boost in the $\bar{x}$ direction. Again 
we fix the energy $p=\gamma M$ and the Kerr parameter $a$ in the boost 
and use the notation $P=\gamma\mu$. By the Lorentz transformation
\begin{align}
\bar{t}&=\gamma(t-vx),\\
\bar{x}&=\gamma(-vt+x),\\
\bar{y}&=y,\\
\bar{z}_i&=z_i,
\end{align}
the metric becomes
\begin{equation}
ds^2=ds_F^2+
\frac{\mu\bar{r}^{7-D}}{\bar{r}^4+a^2Z^2}
\left(d\bar{t}^2+\frac{\bar{r}^2\bar{x}^2+a^2{y}^2-2a\bar{r}\bar{x}{y}}{(\bar{r}^2+a^2)^2}d\bar{x}^2
+\frac{2(\bar{r}\bar{x}-a{y})}{\bar{r}^2+a^2}
d\bar{t}d\bar{x}\right)+O(\gamma^{-1}),
\end{equation}
where $ds_F^2=-dt^2+dx^2+dy^2+\sum_idz_i^2$. In the limit $v\to 1$, this is reduced to
\begin{align}
&~~~~~~~~ds^2=ds_F^2+\lim_{v\to 1}f\left(\gamma(x-vt), y, Z\right)(dt-dx)^2,
\label{transverse_limit_f}\\
&f\left(\gamma(x-vt), y, Z\right)=\frac{\gamma P \bar{r}^{7-D}}{\bar{r}^4+a^2Z^2}
\left(2-\frac{Z^2+(y-a)^2}{\bar{r}^2+a^2}+\frac{2a^2y^2}{\left(\bar{r}^2+a^2\right)^2}\right),
\label{transverse_f}
\end{align}
where we have omitted the
odd terms concerning $\bar{x}$ with the same reason as the parallel case. 

The method for obtaining this limit is the same as the parallel case. We find the 
primitive function $\tilde{\Phi}$ and take the limit as follows:
\begin{equation}
\lim_{v\to 1}f\left(\gamma(x-vt), y, Z\right)
=\frac{\partial}{\partial x}\left\{\lim_{v\to 1}\tilde{\Phi}\left(\gamma(x-vt), y, Z\right)\right\}.
\end{equation}
Again $\tilde{\Phi}$ is reduced to
\begin{equation}
\lim_{v\to 1}\tilde{\Phi}\left(\gamma(x-vt), y, Z\right)=(1/2)\textrm{sgn}(x-t)\Phi(y,Z)+\bar{\Phi}\left(\gamma(x-t),y,Z\right),
\end{equation}
and we obtain
\begin{equation}
\lim_{v\to 1}f=\Phi(y,Z)\delta(x-t)
+\frac{\partial}{\partial x}\bar{\Phi}\left(\gamma(x-t),y,Z\right).
\end{equation}

Now we find the primitive function $\tilde{\Phi}$. 
Using eq.~\eqref{transverse_rbar}, we find the following relations
\begin{align}
&~~~~~~~~~~~\frac{\partial\bar{r}}{\partial x}=\gamma\frac{\bar{r}^3}{\bar{r}^4+a^2Z^2}
\gamma(x-vt),
\\
&\gamma(x-vt)=\textrm{sgn}(x-vt)\sqrt{\bar{r}^2+a^2-y^2-Z^2+\frac{a^2Z^2}{\bar{r}^2}},
\end{align}
with which we derive
\begin{equation}
f=\lim_{v\to 1}\textrm{sgn}(x-vt)
\frac{P\bar{r}^{5-D}}{\sqrt{(\bar{r}^2-\alpha)(\bar{r}^2-\beta)}}
\left(2-\frac{Z^2+(y-a)^2}{\bar{r}^2+a^2}+\frac{2a^2y^2}{(\bar{r}^2+a^2)^2}\right)\frac{\partial\bar{r}}{\partial x},
\end{equation}
where $\alpha$ and $\beta$ are given by
\begin{align}
\alpha&=\frac12\left[y^2+Z^2-a^2+\sqrt{(y^2+Z^2-a^2)^2+4a^2Z^2}\right],\\
\beta&=\frac12\left[y^2+Z^2-a^2-\sqrt{(y^2+Z^2-a^2)^2+4a^2Z^2}\right].
\end{align}
Because $\bar{x}=0$ corresponds to $R\equiv \bar{r}^2=\alpha$, 
we obtain
\begin{equation}
\tilde{\Phi}/P=(1/2)\textrm{sgn}(x-vt)\left[2I^{(D)}_1(R)-\left(Z^2+(y-a)^2\right)I^{(D)}_2(R)+2a^2y^2I^{(D)}_3(R)\right],
\end{equation}
where
\begin{align}
&I^{(D)}_1(R)=\int_{\alpha}^{R}\frac{R^{2-D/2}}{\sqrt{(R-\alpha)(R-\beta)}}dR,
\label{trans_I1}\\
&I^{(D)}_2(R)=\int_{\alpha}^{R}\frac{R^{2-D/2}}{(R+a^2)\sqrt{(R-\alpha)(R-\beta)}}dR,
\label{trans_I2}\\
&I^{(D)}_3(R)=\int_{\alpha}^{R}\frac{R^{2-D/2}}{(R+a^2)^2\sqrt{(R-\alpha)(R-\beta)}}dR.
\label{trans_I3}
\end{align}
Because taking the limit $v\to 1$ for $t\neq z$ corresponds to $R\to \infty$
and each integral is finite at $R=\infty$ for $D\ge 5$,
we find
\begin{align}
&\Phi(X,Y)=2I^{(D)}_1(\infty)-\left[Z^2+(y-a)^2\right]I^{(D)}_2(\infty)+2a^2y^2I^{(D)}_3(\infty),
\label{transverse_F}\\
&\bar{\Phi}\left(\gamma(x-t),X,Y\right)=0.
\end{align}
The function $\Phi$ can be expressed in terms of the elementary functions
for even $D$ and in terms of the elliptic integrals for odd $D$. 
The formulas necessary for this calculation are given in Appendix B.

Now we consider the $D=4$ case before looking at the behavior of $\Phi$.
In this case, $I_1^{(4)}$ is given by
\begin{align}
I_1^{(4)}(R)&=\log\left[2R-\alpha-\beta+2\sqrt{(R-\alpha)(R-\beta)}\right]
-\log\left(\alpha-\beta\right)
\nonumber\\
&\to \log\left|4\gamma^2(x-t)^2\right|-\log\left(\alpha-\beta\right)
~(v\to 1),
\label{transverse_I1(4)}
\end{align}
and we show $I_2^{(4)}(R)$ and $I_3^{(4)}(R)$ in Appendix B.
The limit $\lim_{v\to 1}f$ becomes
\begin{equation}
\lim_{v\to 1}f/P=\frac{2}{|x-t|}+\left\{
-2\log(\alpha-\beta)
-\left[z^2+(y-a)^2\right]I_2^{(4)}(\infty)
+2a^2y^2I_3^{(4)}(\infty)
\right\}\delta(x-t).
\end{equation} 
Although we should
calculate two cases $ay<0$ and $ay>0$ separately
because $I_2(\infty)$ and $I_3(\infty)$ contain $|ay|$, 
they give the same result:
\begin{equation}
\lim_{v\to 1}f=2P\left\{\frac{1}{|x-t|}-\left[\log\left((y-a)^2+z^2\right)+1\right]
\delta(x-t)\right\}.
\end{equation}
This coincides with the previous results of~\cite{BN, BH03}. Again we should mention that
there remains the ambiguity in the form of eq.~\eqref{transverse_I1(4)},
because we can substitue the arbitrary function $g(y,z)$  
as follows:
\begin{equation}
I_1^{(4)}(R)=\log\left\{\left[2R-\alpha-\beta+2\sqrt{(R-\alpha)(R-\beta)}\right]/e^{g(y, z)}\right\}
+g(y, z)-\log\left(\alpha-\beta\right).
\end{equation}
Hence we are required to check the consistency with the
boosted energy-momentum tensor or the boosted Riemann tennsor
and thus the above derivation is just the formal one. 
For the higher-dimensional cases, there is no such ambiguity
because all integrals \eqref{trans_I1}, \eqref{trans_I2} and \eqref{trans_I3} 
are finite for $R\to \infty$.

\begin{figure}[tb]
\centering
{\includegraphics[width=0.35\textwidth]{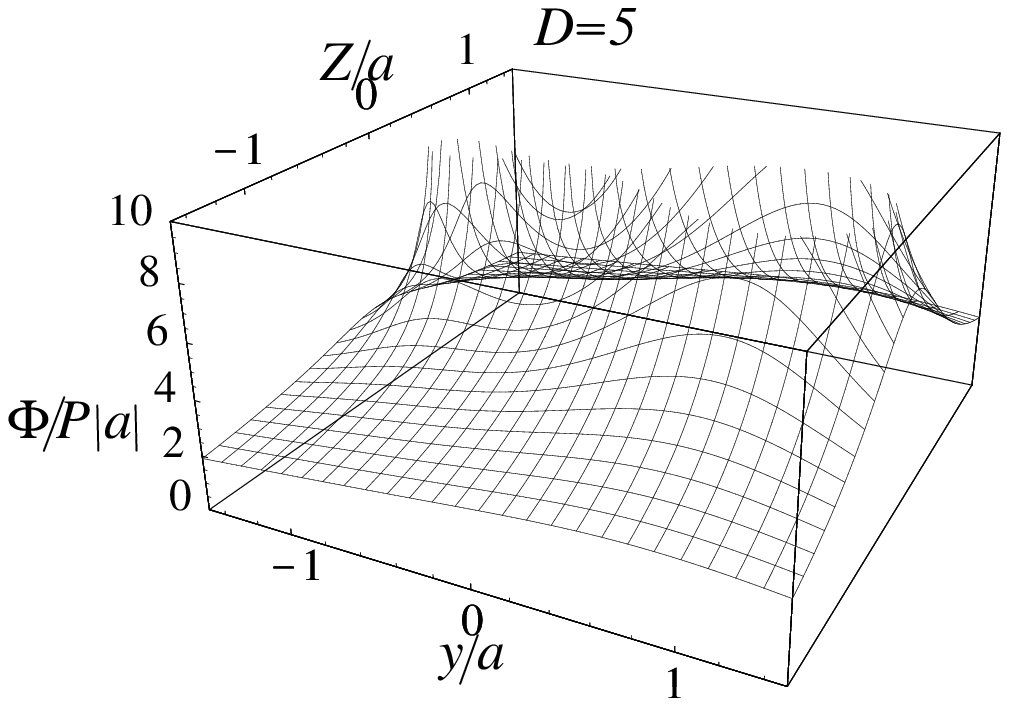}\hspace{5mm}
\includegraphics[width=0.35\textwidth]{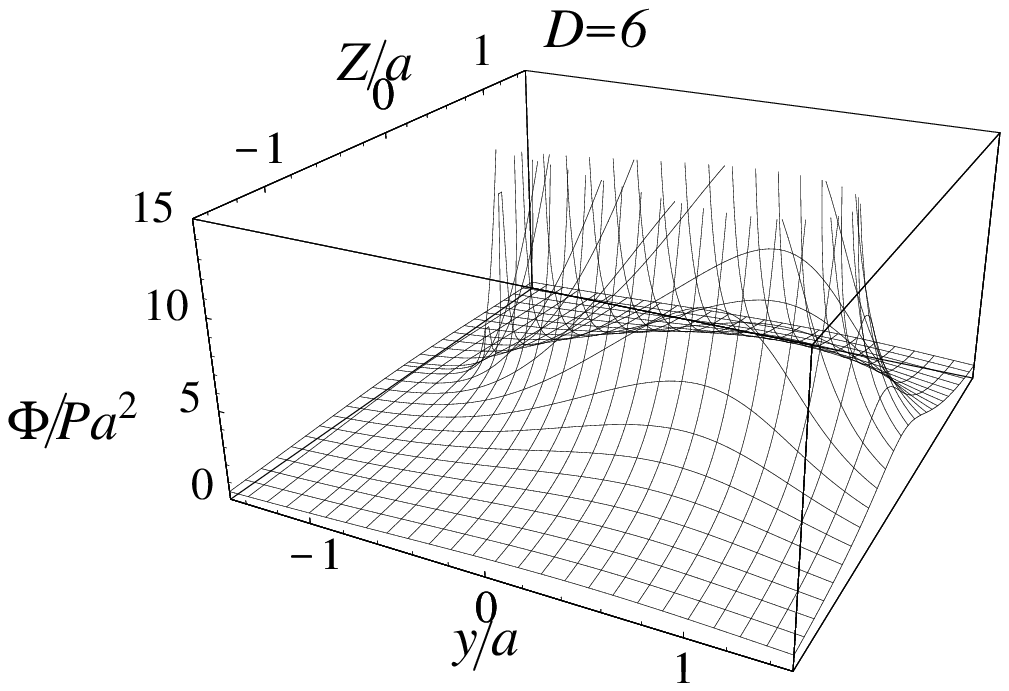}\\
\vspace{5mm}
\includegraphics[width=0.35\textwidth]{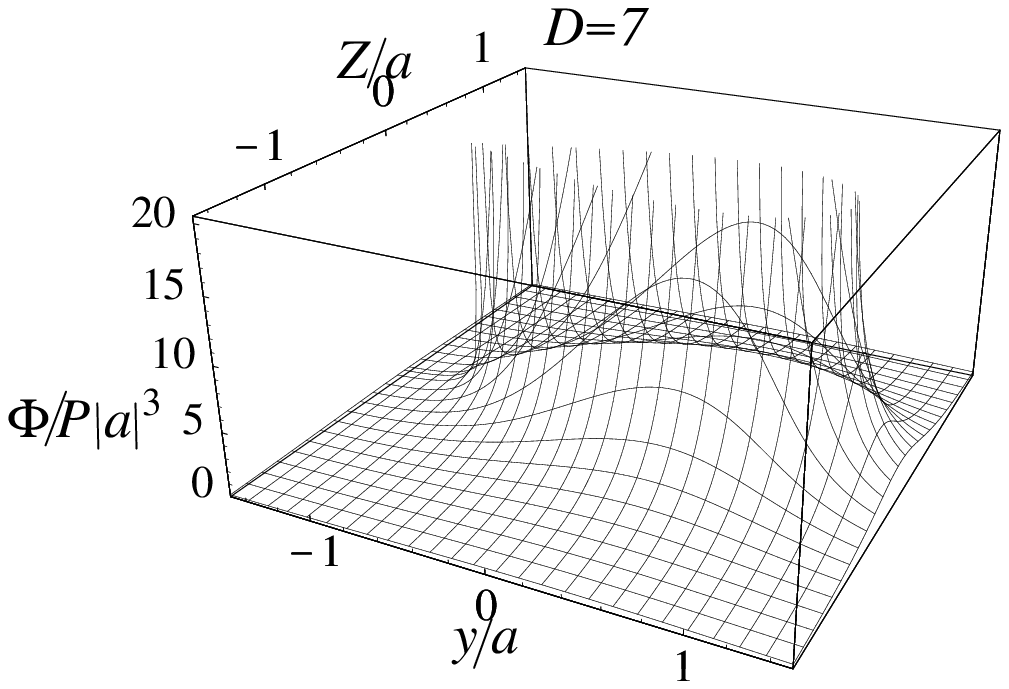}\hspace{5mm}
\includegraphics[width=0.35\textwidth]{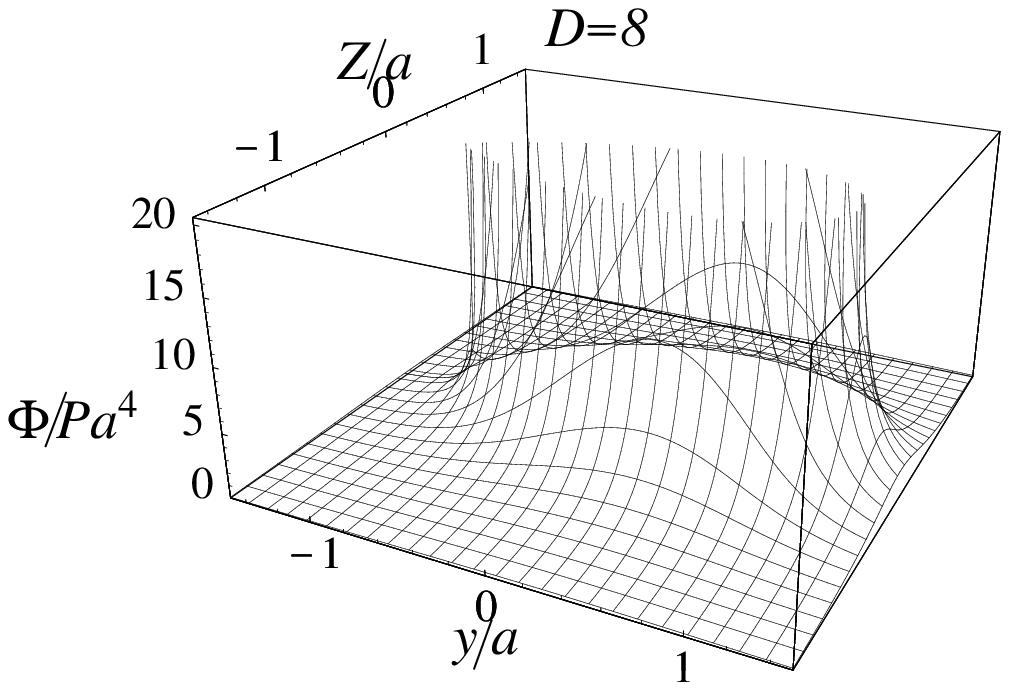}}
\caption{The behavior of the function $\Phi$ for $D=5,...,8$ 
in the transverse case. The unit of $y$
and $Z$ is $a$, and the unit of $\Phi$ is $P|a|^{D-4}$. 
For all $D$, 
$\Phi$ diverges at $Z=0, -a\le y\le a$, which indicates that the 
singularity is a one-dimensional segment.}
\end{figure}

Figure 2 shows the potential $\Phi$ of the shock wave 
for $D=5,...,8$. 
For all $D\ge 5$, the potential diverges at $Z=0, -a\le y\le a$.
Hence in the transverse case,
the singularity has the shape of a one-dimensional segment for all $D\ge 5$.
This is interpreted as follows. In the boost, the ring singularity
becomes naked for $D=5$ and the horizon becomes extremely oblate for $D\ge 6.$
This singularity or the horizon is infinitely Lorentz-contracted in the
direction of the motion. This leads to the segmental shape of the singularity
in the gravitational shock wave. 
The behavior 
of the potential $\Phi$ 
around the singularity is $\sim -\log |Z|$ for $D=5$,
and $\sim 1/|Z|^{D-5}$ for $D\ge 6$. 
Similarly to the parallel case, we see that the relation between the
potential behavior near the singularity and the dimensionality 
of the singularity is the same as the Newtonian gravity.
The behavior of the potential is not symmetric to the $y=0$ plane.
This is the manifestation of the fact that the lightlike Kerr black hole
carries the angular momentum $J=-2ap/(D-2)$. The rotating singularity has
the velocity in the direction of the boost at $(y,Z)=(a,0)$ and
it has the velocity in the converse direction of the boost at $(y,Z)=(-a,0)$. 
Correspondingly, the divergence of the potential at $(y,Z)=(a,0)$ is stronger than
the one at $(y,Z)=(-a,0)$.  

\section{Summary and discussion}

In this paper, we have developed the method for taking the lightlike limit of the boosted
Kerr black holes in the higher-dimensional spacetimes. We analyzed two cases,
the boosts in the parallel direction and in the transverse direction to the spin.
Our method is quite simple: just boosting the metric without 
considering the energy-momentum tensor or the Riemann tensor. 
The problem is reduced to taking the limit $v\to 1$ of the function $f$ of 
eqs.~\eqref{parallel_limit_f} and \eqref{transverse_limit_f}. 
We solved this problem by introducing the primitive function $\tilde{\Phi}$ of $f$,
taking the limit $v\to 1$ of $\tilde{\Phi}$, and then differentiating it. Because
the primitive function $\tilde{\Phi}$ is finite for the higher-dimensional cases,
there is no ambiguity which occurred in the four-dimensional case. 
Hence, we consider that the boost of the metric is the simplest
and easiest method in the higher-dimensional cases.

We discuss whether this method works for the general systems. 
Also in the general systems, the lightlike limit of the metric would have
the form like eq.~\eqref{parallel_limit_f} or eq.~\eqref{transverse_limit_f}. 
The problem is whether we can construct the primitive function of $f$.
If not, we cannot analytically derive the metric of the lightlike system. 
This is the limitation in our method. But even if we cannot find the primitive function of $f$, 
there remains the possibility that we can numerically construct the potential $\Phi$.
Namely, if $\Phi$ is reduced to the form like eqs.~\eqref{para_I1},
\eqref{para_I2}, \eqref{para_I3} and \eqref{parallel_F},
it is easy to calculate the potential $\Phi$ at each point by numerical integrals.
Hence, our analysis shows the direction of the calculation at which we should
aim in general cases.  We also would like to mention that we can guess
the resulting potential $\Phi$ to some extent even if we fail the above 
directions, because the results are quite consistent with our intuition. 
We can intuitively imagine the shape of the infinitely Lorentz-contracted
source and guess the shape of the potential $\Phi$ by the Newtonian analog.

Now we discuss the possible implications of our results in the context
of the brane world scenario. There are several discussions that
the Kerr black hole might provide the gravitational field of 
the elementary particle with spin~\cite{ref3}, especially
motivated by the fact that the Kerr-Newman black hole has
the same gyromagnetic ratio as that of the electron. We consider that this
scenario is unlikely, because the electron in this model has the characteristic
scale $a$, whose value is similar to its Compton wave length.   
Furthermore, we do not have the criterion for determining $a$ in the
case of a massless spinning particle.  
However, we can expect that the spin has an effect on the gravitational
field of the elementary particle, although the spin angular momentum
and the orbital angular momentum are different. This expectation
comes from the fact that the sum of the spin and orbital angular momentum
is the conserved quantity.  It is natural that the conserved quantity affects the
spacetime structure. This expectation is strengthened
by the following observation. If the black hole forms in the head-on collision
of spinning particles,  the produced black hole would have the angular momentum
that corresponds to the sum of two spins. 
In order that the black hole has the angular momentum,
the gravitational field of each particle should not be the same as that of the massless
point particle given in \cite{AS71}.

Because the lightlike Kerr black hole has the quite different gravitational field
from that of the massless point particle, we can also expect that the spin of a particle
has a large effect on its gravitational field. Hence it would affect the condition
of the black hole formation in the high-energy particle collisions.
To clarify this, the investigation of the Einstsein-Dirac system would be useful.
As the first step, the classical field treatment would be appropriate. 
The formalism of this system was  recently developed by Finster~\cite{Fin98}, and
Finster {\it et al.} constructed the spin singlet state 
of two fermions~\cite{FSY99}. 
The generalization to the spin eigenstate of one particle and
its lightlike boost would be an interesting next problem.

Now we turn to the discussion about another possible implication, the relation 
between the Kerr black holes and the strings. 
In the string theory, there are several discussions that some
black holes should be regarded as the elementary particles~\cite{ref4}.
Our formalism would also be applicable to these black hole solutions.
Furthermore, Nishino~\cite{Nis95} and Nishino and Rajpoot~\cite{NR02} 
discussed the fact that the Kerr black hole might be the classical model of the gravitational
field around the closed string,  
simply because the Kerr black hole has the distributional energy-momentum tensor
on its singularity. Hence, our boosted Kerr solution might have an implication
for the black hole formation in the collision of high-energy strings with characteristic
length scale $a$.

Investigatiton of the properties of the 
black hole formation in the collision of two lightlike Kerr strings requires
the analysis of the apparent horizon, which we have not done.
But using the fact that each incoming Kerr string has a singularity with the
characteristic scale $a$, we can discuss the condition for the black hole formation 
as follows. For some impact parameter $b$, 
we write the horizon shape in the case of the two-point-particle collision which
has been calculated in \cite{YN03}. Then we draw on it the shape of the singularity
of the lightlike Kerr strings with Kerr parameter $a$. 
If the singularity crosses the horizon, 
the apparent horizon would not form for such $b$ and $a$. 
Under this assumption, we can calculate the maximal impact
parameter of the black hole production as a function of $a$.   

\begin{figure}[tb]
\centering
{\includegraphics[width=0.45\textwidth]{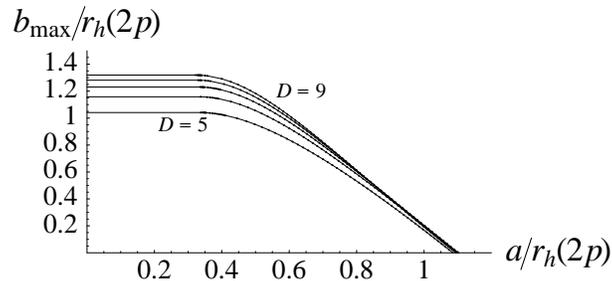}}
\caption{The expected relation between $a$ and the maximal impact parameter 
$b_{\rm max}$ for the black hole production in the collision of two
lightlike Kerr strings boosted in the parrallel direction to the spin.
The unit of axis is $r_h(2p)$.
We see that $b_{\rm max}=0$ in the case of $a\gtrsim r_h(2p)$ for all $D$.}
\end{figure}

Figure 3 shows the expected maximal impact parameter $b_{\rm max}$
for the black hole production 
in the collision of two Kerr strings which are boosted 
in the direction of the spin.
If the Kerr parameter $a$ (i.e., the length of the string) is small compared
to the gravitational radius of the system energy $r_h(2p)$, 
it does not affect the black hole formation.
But $b_{\rm max}$ decreases with the increase in $a$, and
becomes zero at $a\sim r_h(2p)$. 
We can write a similar figure in the case of two lightlike Kerr strings
which are boosted in the transverse direction to the spin.
Hence, the condition for $b_{\rm max}\neq 0$
is written as $a\lesssim r_h(2p)$. 
If the Planck energy is $O({\rm TeV})$ and the string length $a$ is
similar to the Planck length, the gravitational radius for the two particle system
becomes $r_h(2p)\sim a$ if the incoming energy is (few) TeV.
Hence the string length might have the effect such that it prevents the black hole formation.

We should mention that the above discussion is not rigorous, because
we do not know the validity of the assumption we have used.
We can also have the totally different expectation, because
in the higher-dimensional spacetime, the arbitarily long black hole can form
as clarified by Ida and Nakao~\cite{IN02}. The higher-dimensional effect might
lead to the formation of the long apparent horizon in the high-energy
string collision depending on the shape of two strings. 
If this is true, the cross section for the black hole
production might be enhanced. Furthermore, we can also 
expect the formation of the black ring, 
which is the higher-dimensional black hole whose
horizon topology is $S^1\times S^{D-3}$. (See \cite{ER02}
for the five-dimensional solution.) 
If the shape of two strings at the instant of the collision is similar to the ring,
the ring-shaped apparent horizon that surrounds the two strings
might form. Although the author and Nambu discussed
one possible method to produce the black rings in particle systems in the previous paper~\cite{YN04},
it required more than two particles. If the black ring forms due to the
effect of the string length, it would be the more interesting phenomena.  
To clarify the effect of the string length on the black hole formation
in the high-energy string collisions 
would be the interesting problem to be tackled as the next step.

\acknowledgments

The author thanks Ken-ichi Nakao, Yasusada Nambu and Akira Tomimatsu for helpful discussions.
I also thank Kin-ya Oda, Daisuke Ida and Yasunari Kurita for their encouragement.
This work is partly supported by the grant-in-aid for the Nagoya
University 21st Century COE Program (ORIUM).

\appendix

\section{The primitive functions for parallel case}

In this section, we provide the formulas that are necessary for
calculating $I_{i}^{(D)}(\infty)$ for $i=1,2,3$ in the parallel case.
Since the relations
\begin{align}
I_1^{(D)}(R)&=I_3^{(D-2)}(R),\\
I_2^{(D)}(R)&=a^{-2}\left(I_1^{(D)}(R)-I_2^{(D-2)}(R)\right)
\end{align}
hold, all we need are $I_1^{(D)}(\infty)$ for $D=5, 6$, $I_2^{(D)}(\infty)$ for $D=5,6$, and $I_3^{(D)}(\infty)$ for $D=5,6,...$.
These are given in terms of the elliptic integral of the first kind $F(\phi, k)$, of
the second kind $E(\phi, k)$, and of the third kind $\Pi(\phi; c, k)$ which are defined
by
\begin{align}
&F(\phi, k)=\int_0^{\phi}\frac{d\theta}{\sqrt{1-k^2\sin^2\theta}},
\label{elliptic_1}\\
&E(\phi, k)=\int_0^{\phi}\sqrt{1-k^2\sin^2\theta}d\theta,
\label{elliptic_2}\\
&\Pi(\phi; c, k)=\int_0^{\phi}\frac{d\theta}{(1+c\sin^2\theta)\sqrt{1-k^2\sin^2\theta}}.
\label{elliptic_3}
\end{align}
The results are as follows:
\begin{align}
I_1^{(5)}(R)&=\frac{2}{\sqrt{\alpha+a^2}}F(\phi_1, k_1),\\
I_1^{(6)}(R)&=\frac{2}{\sqrt{(\alpha-\beta)a^2}}F(\phi_2, k_2),
\end{align}
\begin{align}
I_2^{(5)}(R)&=\frac{2}{\alpha+a^2}\sqrt{\frac{R-\alpha}{(R-\beta)(R+a^2)}}
+\frac{2}{(\beta+a^2)\sqrt{\alpha+a^2}}\left(F(\phi_1, k_1)-E(\phi_1, k_1)\right),\\
I_2^{(6)}(R)&=\frac{2}{(\alpha+a^2)(\beta+a^2)}\left[-\sqrt{\frac{(R-\alpha)(R-\beta)}{R(R+a^2)}}+\sqrt{\frac{\alpha-\beta}{a^2}}E(\phi_2, k_2)\right]
+\frac{2}{(\alpha+a^2)\sqrt{(\alpha-\beta)a^2}}F(\phi_2, k_2),
\end{align}
\begin{align}
I_3^{(5)}(R)&=\frac{2}{\alpha\beta\sqrt{\alpha+a^2}}
\left[
-(\alpha-\beta)\Pi(\phi_1; -\beta/\alpha, k_1)+\alpha F(\phi_1, k_1)
\right],\\
I_3^{(6)}(R)&=\frac{-2}{\alpha\beta\sqrt{(\alpha-\beta)a^2}}
\left[-\alpha F(\phi_2, k_2)+(\alpha-\beta)E(\phi_2, k_2)\right],
\end{align}
\begin{multline}
I_3^{(7)}(R)=\frac{1}{\alpha a^2R}\sqrt{\frac{(R-\alpha)(R+a^2)}{R-\beta}}
+\frac{[\alpha\beta-a^2(\alpha+\beta)](\alpha-\beta)}{\alpha^2\beta^2a^2\sqrt{\alpha+a^2}}\Pi(\phi_1; -\beta/\alpha, k_1)\\
+\frac{\alpha\beta+a^2(\alpha+\beta)+\beta^2}{\alpha\beta^2(\beta+a^2)\sqrt{\alpha+a^2}}
F(\phi_1, k_1)
-\frac{\sqrt{\alpha+a^2}}{\alpha\beta a^2}E(\phi_1, k_1),
\end{multline}
\begin{multline}
I_3^{(8)}(R)=
\frac{-2}{3\alpha\beta a^2}\sqrt{\frac{(R-\alpha)(R-\beta)(R+a^2)}{R^3}}
+\frac{2(\beta^2+\beta a^2+2\alpha a^2-\alpha\beta)}{3\alpha\beta^2a^2\sqrt{a^2(\alpha-\beta)}}F(\phi_2, k_2)\\
+\frac{4[\alpha\beta-a^2(\alpha+\beta)]}{3\alpha^2\beta^2a^2}\sqrt{\frac{\alpha-\beta}{a^2}}
E(\phi_2, k_2),
\end{multline}
\begin{multline}
I_3^{(9)}(R)=
\frac{\left\{2\alpha\beta a^2-\left[3\alpha\beta-a^2(\alpha+3\beta)\right]R\right\}}{4\alpha^2\beta R^2}\sqrt{\frac{(R-\alpha)(R+a^2)}{R-\beta}}\\
+\frac{[a^2(3\alpha^2+3\beta^2+2\alpha\beta)+\alpha\beta(\alpha-\beta)]}{4\alpha^2\beta^3a^2\sqrt{\alpha+a^2}}F(\phi_1, k_1)
+\frac{3[\alpha\beta-a^2(\alpha+\beta)]\sqrt{\alpha+a^2}}{4\alpha^2\beta^2a^4}
E(\phi_1, k_1)\\
-\frac{\{3[\alpha^2\beta^2+a^4(\alpha^2+\beta^2)]-2\alpha\beta a^2(\alpha+\beta-a^2)\}(\alpha-\beta)}{4\alpha^3\beta^3a^4\sqrt{\alpha+a^2}}
\Pi(\phi_1; -\beta/\alpha, k_1),
\end{multline}
\begin{multline}
I_3^{(10)}(R)=
\frac{2\left\{-3\alpha\beta a^2+4\left[\alpha\beta-a^2(\alpha+\beta)\right]R\right\}}
{15\alpha^2\beta^2 a^4 R^2}
\sqrt{\frac{(R-\alpha)(R-\beta)(R+a^2)}{R}}\\
+
\frac{2\left\{
3\alpha\beta a^2(3\alpha+\beta-3a^2)
+4[\alpha\beta-a^2(\alpha+\beta)](\alpha\beta-\beta a^2-2\alpha a^2-\beta^2)
\right\}}{15\alpha^2\beta^3a^4\sqrt{a^2(\alpha-\beta)}}F(\phi_2, k_2)
\\
-\frac{2\left\{
9\alpha\beta a^2(\alpha+\beta-a^2)
+8[\alpha\beta-a^2(\alpha+\beta)]^2
\right\}}{15\alpha^3\beta^3a^4}
\sqrt{\frac{\alpha-\beta}{a^2}}
E(\phi_2, k_2),
\end{multline}
where $\phi_1, \phi_2, k_1,$ and $k_2$ are given by
\begin{align}
\phi_1&=\arcsin{\sqrt{\frac{R-\alpha}{R-\beta}}}\to \pi/2 ~~(v\to 1),\\
\phi_2&=\arcsin{\sqrt{\frac{a^2(R-\alpha)}{(\alpha+a^2)R}}}\to \arcsin\sqrt{\frac{a^2}{\alpha+a^2}} ~~(v\to 1),
\end{align}
\begin{align}
k_1&=\sqrt{\frac{\beta+a^2}{\alpha+a^2}},\\
k_2&=\sqrt{\frac{-\beta(\alpha+a^2)}{a^2(\alpha-\beta)}}.
\end{align}
Using these formulas, we can easily calculate $I_i^{(D)}(\infty)$ for $i=1,2,3$
and obtain the value of $\lim_{v\to 1}f$ using eq.~\eqref{parallel_F}.

\section{The primitive functions for transverse case}

In this section, we provide the formulas that are necessary for
calculating $I_{i}^{(D)}(\infty)$ for $i=1,2,3$ in the transverse case.
Since the relations
\begin{align}
I_2^{(D)}(R)&=a^{-2}\left(I_1^{(D)}(R)-I_2^{(D-2)}(R)\right),\\
I_3^{(D)}(R)&=a^{-2}\left(I_2^{(D)}(R)-I_3^{(D-2)}(R)\right)
\end{align}
hold, all we need are $I_1^{(D)}(\infty)$ for $D=5, 6,...$, $I_2^{(D)}(\infty)$ for $D=4,5$, and $I_3^{(D)}(\infty)$ for $D=4, 5$. These are given in terms of the elementary functions
for even $D$, while they are given by 
the elliptic integrals~\eqref{elliptic_1}, \eqref{elliptic_2} and \eqref{elliptic_3} and
the complete elliptic integrals $K(k)=F(\pi/2, k)$, 
$E(k)=E(\pi/2, k)$ and $\Pi(c,k)=\Pi(\pi/2; c, k)$ for odd $D$.
The results are as follows:
\begin{align}
I_1^{(5)}(R)&=\frac{2}{\sqrt{\alpha-\beta}}\left(K(k)-F(\phi, k)\right),\\
I_1^{(6)}(R)&=\frac{1}{\sqrt{-\alpha\beta}}\left[
\arctan\left(\frac{-(\alpha+\beta)R+2\alpha\beta}{2\sqrt{-\alpha\beta(R-\alpha)(R-\beta)}}\right)+\frac{\pi}{2}
\right],\\
I_1^{(7)}(R)&=\frac{2}{\beta\sqrt{\alpha-\beta}}
\left[K(k)
-F(\phi, k)+\Pi\left(\phi; \frac{\beta}{\alpha-\beta}, k\right)
-\Pi\left(\frac{\beta}{\alpha-\beta}, k\right)
\right],\\
I_1^{(8)}(R)&=-\frac{\sqrt{(R-\alpha)(R-\beta)}}{\alpha\beta R}
-\frac{2(\alpha+\beta)}{(-\alpha\beta)^{3/2}}
\left[
\arctan\left(\frac{-(\alpha+\beta)R+2\alpha\beta}{2\sqrt{-\alpha\beta(R-\alpha)(R-\beta)}}\right)+\frac{\pi}{2}
\right],
\end{align}
\begin{multline}
I_1^{(9)}(R)=-\frac{2\sqrt{R(R-\alpha)(R-\beta)}}{3\alpha\beta R^2}\\
+\frac{2}{3\alpha\beta^2\sqrt{\alpha-\beta}}
\left\{
(2\alpha+\beta)\left(K(k)-F(\phi, k)\right)
+2(\alpha+\beta)\left[\Pi\left(\phi; \frac{\beta}{\alpha-\beta}, k\right)
-\Pi\left(\frac{\beta}{\alpha-\beta}, k\right)\right]
\right\},
\end{multline}
\begin{multline}
I_1^{(10)}(R)=-\frac{\left[3(\alpha+\beta)R+2\alpha\beta\right]\sqrt{(R-\alpha)(R-\beta)}}{4\alpha^2\beta^2R^2}\\
+\frac{(3\alpha^2+2\alpha\beta+3\beta^2)}{8(-\alpha\beta)^{5/2}}
\left[\arctan\left(\frac{-(\alpha+\beta)R+2\alpha\beta}{2\sqrt{-\alpha\beta(R-\alpha)(R-\beta)}}\right)+\frac{\pi}{2}\right],
\end{multline}
\begin{align}
I_2^{(4)}(R)&=\frac{-1}{\sqrt{(\alpha+a^2)(\beta+a^2)}}
\log\left|\frac{2\sqrt{(\alpha+a^2)(\beta+a^2)(R-\alpha)(R-\beta)}-(\alpha+\beta+2a^2)R+(\alpha+\beta)a^2+2\alpha\beta}{(\alpha-\beta)(R+a^2)}\right|,
\\
I_2^{(5)}(R)&=\frac{2}{(\beta+a^2)\sqrt{\alpha-\beta}}\left[
K(k)-F(\phi, k)+\Pi\left(\phi; \frac{\beta+a^2}{\alpha-\beta}, k\right)
-\Pi\left(\frac{\beta+a^2}{\alpha-\beta}, k\right)
\right],
\end{align}
\begin{equation}
I_3^{(4)}(R)=\frac{1}{2(\alpha+a^2)(\beta+a^2)}
\left[-\frac{2\sqrt{(R-\alpha)(R-\beta)}}{(R+a^2)}
+(2a^2+\alpha+\beta)I_2^{(4)}(R)\right],
\end{equation}
\begin{multline}
I_3^{(5)}(R)=\frac{-1}{a^2(\alpha+a^2)(\beta+a^2)}\left\{
\frac{(\beta+a^2)\sqrt{R(R-\alpha)(R-\beta)}}{(R-\beta)(R+a^2)}
+\sqrt{\alpha-\beta}\left(E(\phi, k)-E(k)\right)\right.\\
+\frac{a^2(2a^2+\alpha+\beta)}{(\beta+a^2)\sqrt{\alpha-\beta}}
\left(F(\phi, k)-K(k)\right)
-\frac{3a^4+2a^2(\alpha+\beta)+\alpha\beta}{(\beta+a^2)\sqrt{\alpha-\beta}}
\left[
\Pi\left(\phi; \frac{\beta+a^2}{\alpha-\beta}, k\right)
-\Pi\left(\frac{\beta+a^2}{\alpha-\beta}, k\right)
\right]
\Bigg\},
\end{multline}
where $\phi$ and $k$ are given by
\begin{equation}
\phi=\arcsin{\sqrt{\frac{\alpha-\beta}{R-\beta}}}\to 0 ~~(v\to 1),
\end{equation}
\begin{equation}
k=\sqrt{\frac{-\beta}{\alpha-\beta}}.
\end{equation}
Using these formulas, we can easily calculate $I_i^{(D)}(\infty)$ for $i=1,2,3$
and obtain the value of $\lim_{v\to 1}f$ using eq.~\eqref{transverse_F}.
Although the resulting potential $\Phi(y, Z)$ is quite complicated in most cases, 
we found a rather simple formula only in the $D=6$ case:
\begin{equation}
\frac{\Phi(y,Z)}{Pa^2}=
2+\frac{(y+a)^2-Z^2}{|aZ|}
\left[\arctan\left(\frac{2|aZ|}{y^2+Z^2-a^2}\right)+\frac{\pi}{2}\right]
+\frac{2(y+a)}{a}\log\left(\frac{(y-a)^2+Z^2}{(y+a)^2+Z^2}\right).
\end{equation}

\end{document}